\documentstyle[12pt,aasms4,flushrt]{article}

\def\myputfigure#1#2#3#4#5%
{\vskip#5pt\makebox[0pt]{\hskip#2in
\includegraphics[width=#3\textwidth]{#1}}\vskip#4pt\hfill}

\newcommand\lsim{\mathrel{\rlap{\lower4pt\hbox{\hskip1pt$\sim$}}
        \raise1pt\hbox{$<$}}}
\newcommand\gsim{\mathrel{\rlap{\lower4pt\hbox{\hskip1pt$\sim$}}
        \raise1pt\hbox{$>$}}}

\begin{document}
\Large 
\centerline{\bf Astronomy: Trouble at first light}
\normalsize 
\author{Piero Madau}

\begin{verbatim}
          News and Views, Nature 440, 1002-1003 (20 April 2006)
\end{verbatim}

\vskip 0.2in 

\hrule 

\vskip 0.1in 

\noindent
{\em\large 
The question of how much light the first stars produced is fundamental to models
of the Universe's development. But observations have so far failed to agree: 
is the answer a lot, or not very much at all?
}

\vskip 0.2in 

\hrule 

\vskip 0.2in 

\thispagestyle{empty}

\noindent 
On page 1018 of this issue, Aharonian et al.\cite{aharonian} report the detection 
of copious high-energy $\gamma$-ray emission from two `blazars' $-$ a class of 
active galaxy $-$ around 2 billion light years from Earth. This observation 
indicates that such radiation can travel largely unimpeded through the cosmos, 
and implies that the infrared glow of the first stars in the Universe and 
their remnants is fainter than previous measurements had led us to believe. If 
true, that could influence our ideas of how and when the first structures 
in the Universe evolved.

\noindent
The formation of structure in the Universe is believed to proceed hierarchically, 
with smaller galaxies merging, through the action of gravity, to build more 
massive ones. But the timing and sequence of the events through which the very 
first galaxies and stars formed remain largely unknown. According to current 
theories, the first dwarf galaxies hosted metal-free stars over a hundred times 
more massive than the Sun. These stars shone intensely for only a few million years 
and then either blew themselves apart in gigantic supernova explosions, or 
collapsed to form the first massive black holes.

\noindent 
Astronomers have long been rummaging through the Universe for tell-tale signs of 
these dramatic beginnings. When the first stars ignited, they emitted large 
numbers of photons at ultraviolet wavelengths. These photons `reionized' the 
surrounding atomic hydrogen gas that had formed as the Universe cooled. Just 
last month, astronomers using NASA's Wilkinson Microwave Anisotropy Probe (WMAP) 
reported the latest detection of photons produced soon after the Big Bang. Their 
data show that these `cosmic microwave background' photons became polarized 
(tending to oscillate in only one direction perpendicular to their line of 
travel) by scattering on free electrons in the early Universe. The level of 
polarization allows the era of reionization to be pinpointed to some 400 
million years after the Big Bang, when the Universe was just 3\% of its present 
age\cite{spergel}.

\noindent
So how much of the background light that we see comes from the first stars? As 
the Universe aged and expanded, part of the ultraviolet radiation emitted by 
these stars was absorbed again by re-formed atomic hydrogen. Lower-energy 
ultraviolet light escaped this fate, but was stretched to longer, redder 
wavelengths. Therefore, although the early stellar populations were twinkling so 
long ago that current telescopes cannot detect them, their combined energy 
output is recorded in diffuse light that reaches Earth in the near-infrared region 
of the electromagnetic spectrum, at wavelengths of a few micrometres. Resolving 
this infrared glow is, however, a daunting task, because many other celestial 
sources $-$ among them older stars in closer galaxies, active galactic nuclei 
known as quasars, and the bright foreground sources in the Milky Way and the 
Solar System $-$ emit radiation at similar wavelengths.

\noindent
Nevertheless, several groups have claimed to have found the footprints of baby 
galaxies at near-infrared wavelengths, using data from NASA's Cosmic Background 
Explorer (COBE)\cite{dwek}$^,$\cite{gorjian}$^,$\cite{cambresy} and Spitzer 
Space Telescope\cite{kashlinsky}, and Japan's Infrared Telescope in Space 
(IRTS)\cite{matsumoto}. Their evidence comes in two forms. First, there is an 
excess signal above the combined emission of normal foreground galaxies that 
would require energetic events to have occurred in the early Universe. Second, 
the very uneven distribution of the radiation could arise from the spatial 
clustering properties of primordial stellar systems.

\noindent
But rather than helping to decipher the epoch of cosmic first light, such 
observations have in fact created another puzzle. Simply stated, the dawn of 
galaxies seems to be too brilliant: the excess signal outshines the cumulative 
emission from all galaxies between Earth and the extremely distant first stars. 
If primordial sources are to account for all of this infrared radiation, current 
models of star formation in the young Universe look distinctly shaky. Too many 
massive stars ending their brief lives in a giant thermonuclear explosion would, 
for instance, eject large amounts of heavy elements such as carbon and oxygen 
into space, polluting the cosmos very early on and altering for ever the 
composition of the raw material available for second-generation stars. But if the 
first-generation stars were to collapse to massive black holes instead, 
gas accretion onto such black holes would produce large amounts of X-rays. 
Both variants seem to be in conflict with current observations\cite{santos}$^,
$\cite{madau}$^,$\cite{salvaterra}.

\noindent Enter Aharonian and colleagues\cite{aharonian}, and their measurements 
of teraelectronvolt (TeV) $\gamma$-ray photons from blazars. These photons, which 
carry $10^{12}$ times more energy than visible light, interact with near-infrared 
photons through the quantum-mechanical process of electron-positron pair 
creation. Through this process, most of the TeV photons are absorbed long 
before they reach Earth. The observed level of $\gamma$-ray attenuation 
can, therefore, 
be used to estimate indirectly the energy density of infrared starlight 
present in intergalactic space.

\noindent
The authors' observations were made with the High Energy Stereoscopic System, 
HESS, inaugurated in Namibia in 2004 and operated by a collaboration of 
scientists from nine countries. HESS uses four large telescopes (Fig. 1) 
arranged at the corners of a 120-metre square to detect the faint flashes, 
lasting only a few billionths of a second, of blue `Cherenkov' light emitted 
when a high-energy $\gamma$-ray hits the atmosphere. Up to four images 
are combined to determine the direction of the $\gamma$-ray and the energy 
it deposits in the atmosphere.

\begin{figure}
\epsscale{1.0}
\plotone{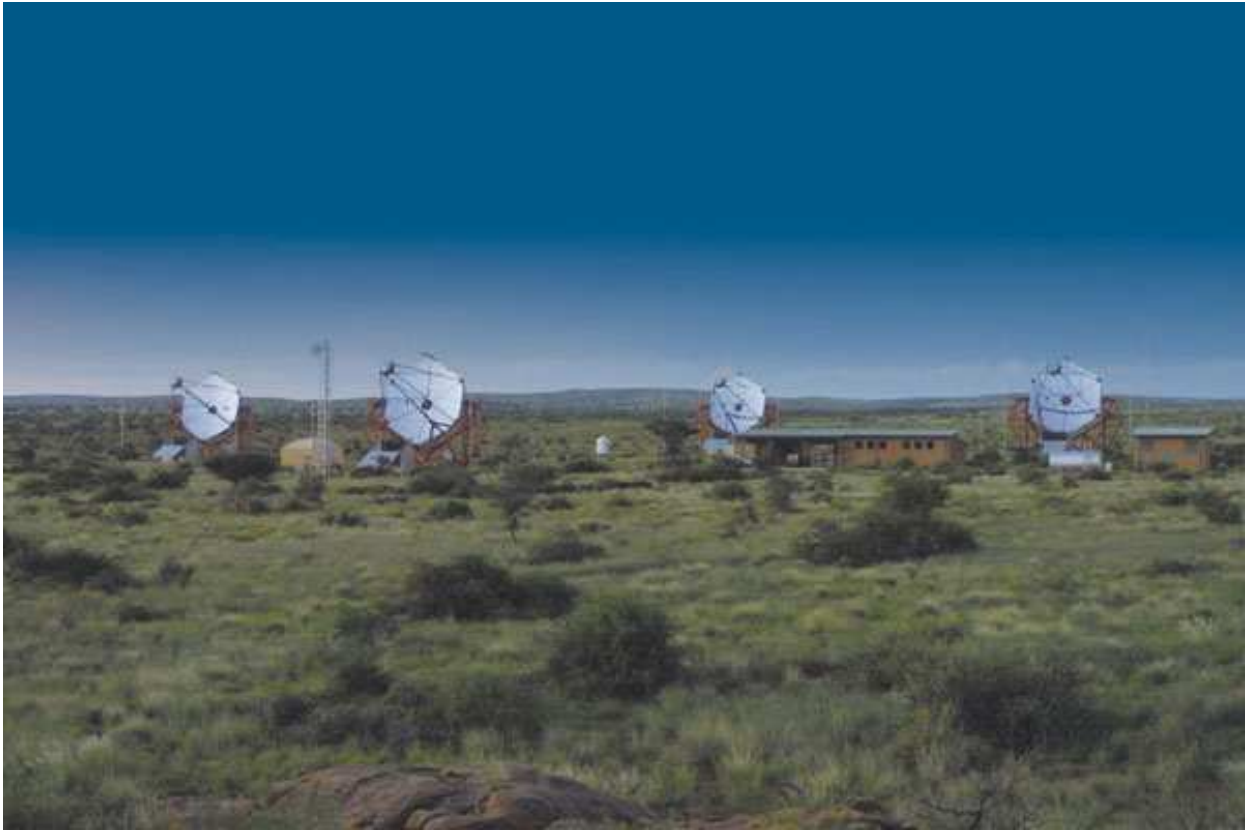}
\caption{Eyes to the light: the four HESS telescopes in the veldt of central Namibia.
}
\end{figure}

\noindent Since June 2004, HESS has accumulated more than 80 hours of 
observations on two blazars. The $\gamma$-rays emitted by these most distant 
known sources have energies between 0.2 and 3 TeV; infrared radiation at 
wavelengths longer than 1 micrometre absorbs $\gamma$-ray photons of energy 
greater than 0.7 TeV. So, if such $\gamma$-rays were propagating through a 
dense sea of infrared photons, as implied by previous measurements\cite{dwek}$^,$
\cite{gorjian}$^,$\cite{cambresy}$^,$\cite{kashlinsky}$^,$\cite{matsumoto},
the spectra of the two blazars recorded by HESS would reveal evidence of strong, 
energy-dependent attenuation. But Aharonian et al.\cite{aharonian} show that 
intergalactic space is more transparent to $\gamma$-rays than would be 
expected if an infrared background excess existed. Remarkably, the attenuation 
in the HESS images seems consistent merely with the integrated infrared 
output from resolved foreground galaxies, together with the total extragalactic 
light produced by second-generation stars according to recent 
theoretical calculations\cite{primack}.

\noindent
What is the cause of the discrepancy between the HESS data and previous results? 
Aharonian and colleagues point out that their interpretation of the HESS 
results relies on the assumption that the intrinsic $\gamma$-ray spectra of 
the two active galaxies are not at odds with current models of blazar behaviour. 
Further high-energy observations of blazars at different cosmological distances 
should settle this issue. But it seems unlikely that they will be able to 
supply the fine-tuning required to make the HESS data consistent with the 
previously reported excess of infrared background light.

\noindent
The distribution of this excess infrared component over different wavelengths 
is almost identical to that of sunlight reflected from local interplanetary 
dust clouds\cite{arendt}. It is therefore conceivable that not all of the 
foreground emission has been subtracted from the diffuse-sky maps, and the excess 
may not be extragalactic after all. On the other hand, the fluctuations detected 
in highly sensitive images taken with the Infrared Array Camera onboard 
Spitzer\cite{kashlinsky} do not change between observations performed six months 
apart, and changes would be expected if their origin was zodiacal light 
within the Solar System.

\noindent 
Whatever the final resolution of the mystery concerning the signature of the 
first stars, the HESS findings\cite{aharonian} will stir much debate 
among cosmologists. They are likely to spark further attempts to glimpse 
the crucial early stages of the galaxy formation process.

\small

\noindent{\it Piero Madau is in the Department of Astronomy and Astrophysics,\\ 
University of California Santa Cruz, CA 95064, USA. \\email: pmadau@ucolick.org}
\end{document}